\documentclass{article}

\usepackage[final, nonatbib]{neurips_2025_ml4ps}

\usepackage[utf8]{inputenc} % allow utf-8 input
\usepackage[T1]{fontenc}    % use 8-bit T1 fonts
\usepackage{hyperref}       % hyperlinks
\usepackage{url}            % simple URL typesetting
\usepackage{booktabs}       % professional-quality tables
\usepackage{amsfonts}       % blackboard math symbols
\usepackage{nicefrac}       % compact symbols for 1/2, etc.
\usepackage{microtype}      % microtypography
\usepackage{xcolor}         % colors
\usepackage{todonotes}         % colors
\usepackage{wrapfig} % in your preamble
\usepackage{amsmath}
 \usepackage[backend=biber,sorting=none,style=numeric]{biblatex}
\title{Evaluation of Novel Fast Machine Learning Algorithms for Knowledge-Distillation-Based Anomaly Detection at CMS}

\author{%
  Lino Gerlach$^*$\\
  Princeton University\\
  \texttt{lg0508@princeton.edu} \\
  \And
  Elliott Kauffman$^*$\\
  Princeton University \\
  \texttt{ek8842@princeton.edu} \\
  \And
  Abhishikth Mallampalli$^*$\\
  University of Wisconsin-Madison \\
  \texttt{amallampalli@wisc.edu} \\
}

\begin{document}

\maketitle
\def\thefootnote{*}\footnotetext{Equal contribution.}
% \def\thefootnote{\arabic{footnote}}
% text text text\footnote{normal footnote}

\begin{abstract}
% The CICADA (Calorimeter Image Convolutional Anomaly Detection Algorithm) project aims to detect anomalous physics signatures without bias from theoretical models in proton-proton collisions at the Compact Muon Solenoid (CMS) experiment at the Large Hadron Collider. CICADA identifies anomalies in low-level calorimeter trigger data using a convolutional autoencoder, whose behavior is transferred to a compact model via knowledge distillation. Careful quantization allows the model to meet the requirement of sub-200ns inference times on FPGAs. 
% While current implementations use Quantization Aware Training with per-layer quantization schemes, a gradient-based quantization optimization method introduced in 2024 enables fine-grained, parameter-level quantization. We present a novel CICADA implementation leveraging this highly granular quantization. Evaluated on CMS open data, it achieves comparable anomaly detection performance with significantly reduced resource usage in emulated FPGA conditions. We also demonstrate the feasibility of analyzing calorimeter energy deposits at finer spatial granularity while maintaining CMS’s stringent latency and hardware constraints for real-time data selection.
% \todo{Copy-pasted the FastML abstract here. Need to adjust}

The CICADA (Calorimeter Image Convolutional Anomaly Detection Algorithm) project aims to detect anomalous physics signatures without bias from theoretical models in proton–proton collisions at the CMS experiment at the Large Hadron Collider. CICADA identifies anomalies in low-level calorimeter trigger data using a convolutional autoencoder, whose behavior is transferred to compact student models via knowledge distillation. Careful model design and quantization ensure sub-200 ns inference times on FPGAs.
We investigate novel student model architectures that employ differentiable relaxations to enable extremely fast inference at the cost of slower training—a welcome tradeoff in the knowledge distillation context. Evaluated on CMS open data and under emulated FPGA conditions, these models achieve comparable anomaly detection performance to classically quantized baselines with significantly reduced resource usage. The savings in resource usage enable the possibility to look at a richer input granularity.

  % The abstract paragraph should be indented \nicefrac{1}{2}~inch (3~picas) on
  % both the left- and right-hand margins. Use 10~point type, with a vertical
  % spacing (leading) of 11~points.  The word \textbf{Abstract} must be centered,
  % bold, and in point size 12. Two line spaces precede the abstract. The abstract
  % must be limited to one paragraph.
\end{abstract}

\section{Introduction}
At CERN's Large Hadron Collider (LHC), tens of millions of proton-proton collisions are produced per second. 
The Compact Muon Solenoid (CMS) detector at the LHC captures complex high-dimensional data from each collision.
With such large data volume and rate, it is imperative to have an effective data reduction strategy.
The Level-1 Trigger (L1T) in the CMS experiment is the first data reduction step, bringing the event rate from 40 MHz to 100 kHz --- a 99.75\% reduction within 3.8 microseconds \cite{trigger_tdr}\cite{trigger_upgrade_tdr}.
The L1T is comprised of real-time algorithms built on Field Programmable Gate Arrays (FPGAs).
%A traditional CMS L1T algorithm selects events based on the expected physics topology of various analysis targets, for example selecting for one electron over a specified energy threshold.
%This strategy is inherently model-dependent, so any search for new physics is thus dependent on physicists predicting the correct event signature to select for.
%To broaden the search for new physics to more possibilities, CMS has implemented anomaly detection-based algorithms in the L1T.
Traditional physics-process oriented triggers are highly efficient for specific, known signal signatures but are inherently inflexible, rendering them ineffective for broad searches for unknown phenomena. Conversely, more general physics object-oriented triggers capture a wider variety of signals but must impose high momentum thresholds to control data rates, severely limiting their acceptance for new physics involving low-energy particles. To broaden the search for new physics to more possibilities, CMS has implemented two Machine Learning (ML)-based anomaly detection algorithms in the L1T: ``Calorimeter Image Convolutional Anomaly Detection Algorithm'' (CICADA) \cite{CMS-DP-2024-121} and ``Anomaly Extraction Online L1 Trigger Lightweight'' (AXOL1TL) \cite{CMS-DP-2024-059}. These new triggers aim to provide an optimal tradeoff by having a low data rate while maintaining a wider reach, all within the demanding, low-latency environment of the L1T, see Fig.~\ref{fig:latency_req}.
\\\\
Our contribution focuses on CICADA, which deploys a convolutional autoencoder that is trained on CMS Zero Bias data, which is a random trigger that selects events with no bias towards any physics signature.

\subsection{Our contribution} This paper explores three different quantization schemes that aim to reduce the resources required to run the CICADA algorithm without sacrificing performance. We show that two novel quantization libraries (HGQ, LGN) offer a vastly superior trade-off between physics performance and resource efficiency compared to the established QKeras baseline. This resource saving directly benefits the LHC physics program by enabling the deployment of more complex real-time algorithms --- for example, analyzing calorimeter energy deposits at finer spatial granularity than previously possible --- while meeting CMS’s stringent real-time data selection constraints.
We also utilize CMS Open Data to ensure the project’s reach to the broader community \cite{CERNOpenData2017}.

\section{Background}

\begin{figure}[htb]
    \centering
    \begin{minipage}{0.35\textwidth}
        \centering
        \includegraphics[width=\linewidth]{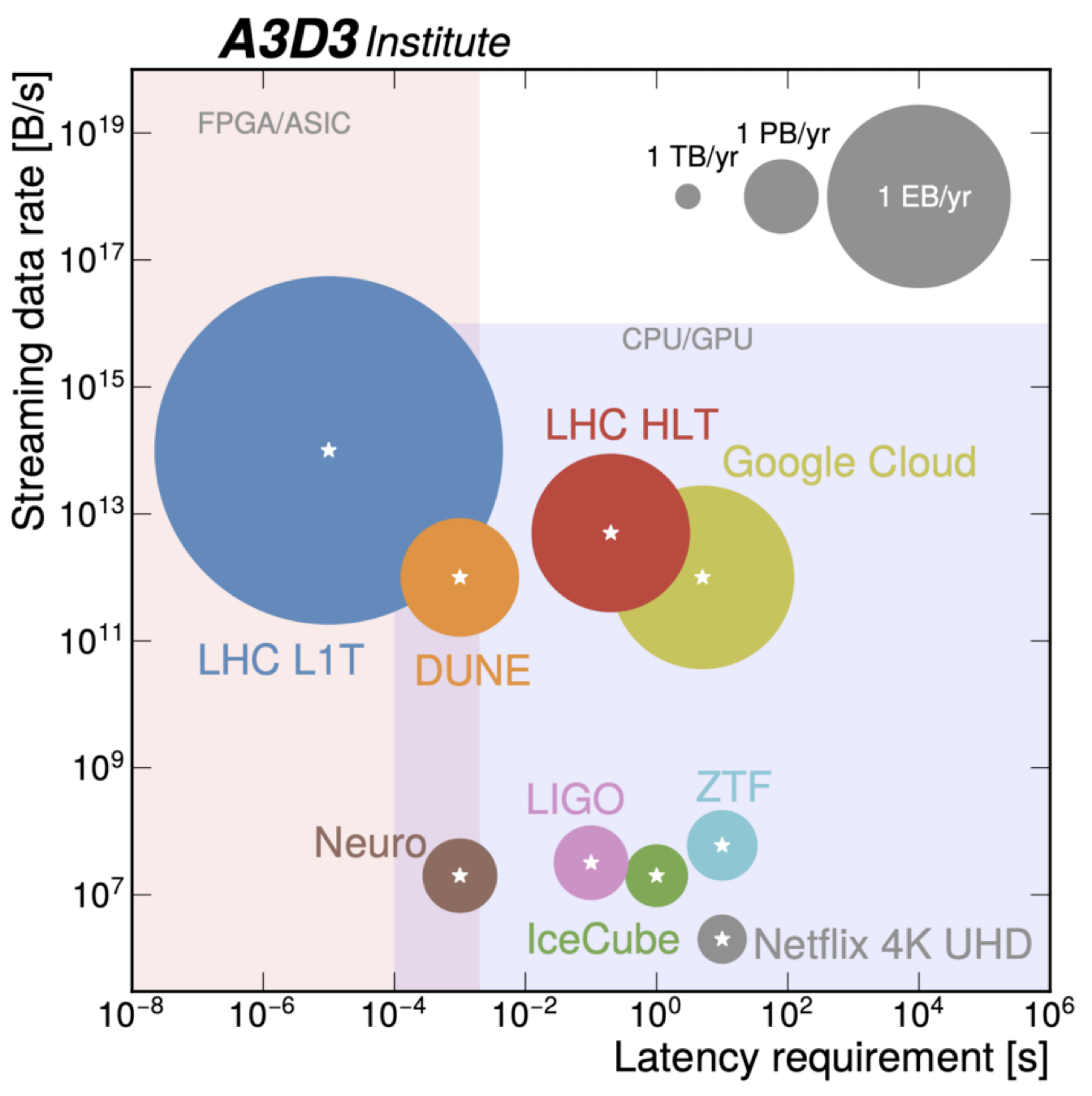}
        \caption{Data rate vs. latency requirements across various scientific and industrial applications.}
        \label{fig:latency_req}
    \end{minipage}%
    \hfill % This command adds flexible space between the two minipages
    \begin{minipage}{0.6\textwidth}
        \centering
        \includegraphics[width=\linewidth]{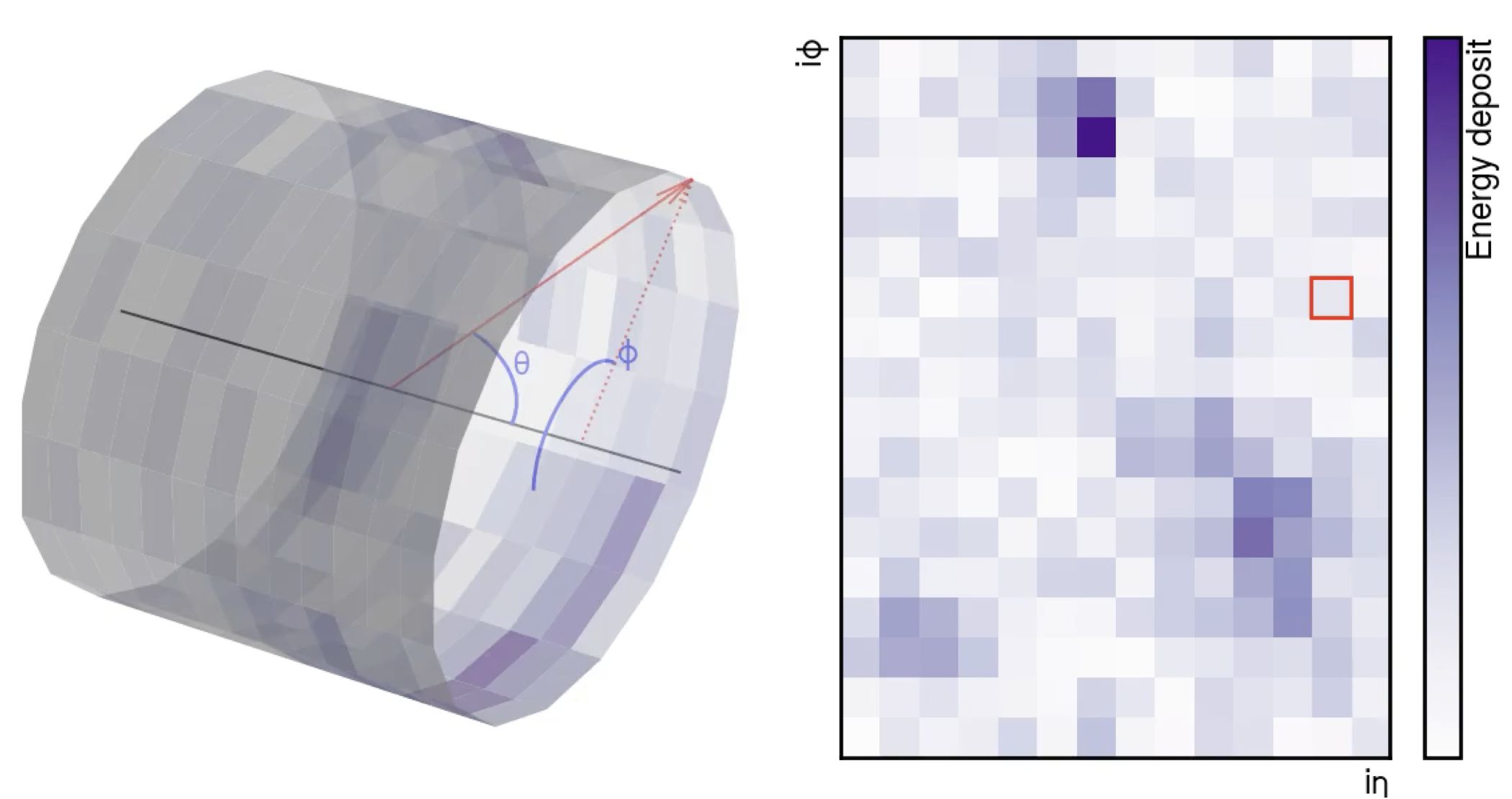}
        \caption{Schematic depiction of the topology of CICADA's inputs. The energy deposits on the unrolled cylindrical calorimeter are image-like data.}
        \label{fig:cicada_input}
    \end{minipage}
\end{figure}

The input to the autoencoder is an $18\phi\times14\eta$ unrolled cylindrical grid of calorimeter energy deposits, as shown in Fig.~\ref{fig:cicada_input}.
Here, $\phi$ refers to the azimuthal angle with respect to the proton beamline and $\eta$ refers to the a pseudorapidity (effecitvely the polar angle). The reconstruction loss is defined as the mean squared error (MSE) of the transverse energy deposits ($E_T$) in the input and output $\phi\times\eta$ grid:
\begin{equation}
    \text{loss} = \frac{1}{252}\sum_{i=1}^{252}\left(E_{T,i}^{\text{original}}-E_{T,i}^{\text{reconstructed}}\right)^2
\end{equation}
This value is used to define the anomaly score, based on the principle that the autoencoder will reconstruct rare physics processes less accurately than the bulk of Zero Bias data it was trained on.

This autoencoder model consists of around 300k parameters and therefore cannot be deployed within the FPGA-based L1T. 
To solve this problem, CICADA employs a knowledge distillation strategy, in which the autoencoder is treated as a ``teacher" model and a much smaller, $\sim$10k parameter convolutional neural network acts as a ``student" \cite{pol2023knowledgedistillationanomalydetection}.
The student model is trained to learn a soft target from the teacher model (the anomaly score) from the original $18\phi\times14\eta$ input, using the Mean Absolute Error (MAE) loss function. 
The soft target is defined from the teacher reconstruction loss as:
\begin{equation}
    Q_{log} := \text{quantize}\left(\log\left(32 \cdot \text{loss} \right)\right)
\end{equation}

This transformation of the loss reduces the range of the targets to assist in training and makes use of the fixed bit-width requirement of the trigger system.
Since the teacher model is trained on Zero Bias data and thus will result in only low anomaly scores, the student model is exposed to events with a known high anomaly score so that it can sufficiently learn the characteristics of high-score events.
This process is known as \emph{outlier exposure}.
To optimize for low-precision inference on hardware, the student model was trained with quantization-aware training (QAT) using QKeras \cite{qkeras}.
The trained QKeras student model was then translated into synthesizable C++ code with hls4ml. hls4ml inserts pragmas into the code to optimize the high-level synthesis \cite{hlsfml}. 
The C++ code was further fine-tuned in order to meet the resource and latency requirements imposed by the trigger system.

\section{Methods and Experimental Setup}

\subsection{Quantization Schemes}

\textbf{HGQ} The High Granularity Quantization (HGQ) library~\cite{HGQ} offers a more fine-grained approach to model compression than QKeras, which operates at the layer level. 
HGQ achieves this sub-layer granularity by treating the bitwidths of individual weights and biases as trainable parameters. 
To optimize these bitwidths, HGQ employs a two-pronged strategy within the loss function. 
First, it intentionally minimizes a proxy for hardware cost by introducing a term called Effective Bit Operations (EBOPs), which is proportional to the number of bitwise operations in the network. 
Second, it adds an L1 regularization penalty directly to the bitwidth parameters, which explicitly pushes them towards lower values. 
The EBOPs and L1 terms are weighted by the hyperparameters  $\beta$ and $\gamma$, respectively, controlling the trade-off between model accuracy and resource efficiency. 
By minimizing this combined loss, the training process simultaneously learns the optimal value for each weight and its ideal bitwidth, automatically pruning connections by learning a bitwidth of zero.

\textbf{LGN} Another strategy for fast and small CICADA student models is by changing the architecture to a logic gate network (LGN). 
LGNs are comprised of binary logic gates accepting only two inputs, so they can execute very quickly. 
To enable gradient-based training, differentiable functions are used as substitutes for the logic gates. 
Upon inference, each neuron is collapsed to the logic gate with the highest probability \cite{petersen2022deepdifferentiablelogicgate}\cite{10.5555/3737916.3741767}, leading to multiplication-free and extremely efficient deployment on hardware.

\subsection{Experimental design}

\paragraph{Dataset} Our study is conducted entirely using public CMS Open Data to ensure reproducibility and enable further community-driven research. 
The model is trained on the 2017 Zero Bias proton-proton collision dataset, which serves as the background distribution. 
As a signal sample, we use top quark pair production ($t\bar{t}$), which has three possible final states characterized by number of leptons ($l$), neutrinos ($\nu$), and quarks ($q$). 
The $t\bar{t}\rightarrow2l+2\nu$ sample is split into three sets: two are used for outlier exposure during training and validation, while the third is reserved for the final evaluation. 
The other two samples, $t\bar{t}\rightarrow l+\nu+2q$ and $t\bar{t}\rightarrow4q$, are used exclusively for evaluation. 
These processes were chosen as they are the only relevant physics signals available in the open data collection.
Additionally, $t\bar{t}$ processes have a much lower cross-section than the majority of processes present in the Zero Bias dataset, meaning that they happen at a comparatively negligible rate, so that they should be anomalous with respect to the majority of events.

\textbf{Experiments} 
To explore the trade-off between performance and resource usage, we performed a hyperparameter scan for both HGQ and LGN. 
For HGQ, multiple student models were trained by varying the \(\beta\) parameter in the loss function, while $\gamma$ remained at the default value, 1e-8.
For LGN, multiple student models with different architectures as well as different bit representations for the input data were trained.

\textbf{Performance} was evaluated using a number of metrics. First, the teacher and student model scores were compared using the Earth Mover’s Distance (EMD) to assess how well the student model learned the target teacher anomaly scores. 
Then, the physics performance was quantified using the Receiver Operating Characteristic Area Under the Curve (ROC AUC) to measure the trade-off between signal efficiency and background rejection. It is worth noting that a typical CMS L1T algorithm will operate at very low false positive rates (<< 1 \textperthousand) to achieve the required data rate reduction. 

\textbf{Hardware Cost} was estimated by synthesizing each model onto the target FPGA platform, an AMD Virtex-7 (XC7VX690T) FPGA, using Vitis HLS~\cite{XilinxVivado2021,XilinxVitisHLS2021}.
The synthesis reports provide key implementation metrics: resource utilization, measured in Look-Up Tables (LUTs), Flip-Flops (FFs), and Digital Signal Processors (DSPs), as well as the processing latency, expressed in the number of FPGA clock cycles required for a single inference.
For the HGQ model, the hls4ml framework was used to generate HLS-compatible code.
For the LGN, the structure was directly translared into C code for HLS synthesis. This was, however, only achieved for dense architectures. Synthesis for convolutional LGNs is work in progress.

% \newpage
\pagebreak

\section{Results}

\begin{wrapfigure}{r}{0.42\textwidth} % 'r' = right, 'l' = left
    \vspace{-20pt} % adjust vertical position
    \centering
    \includegraphics[width=0.45\textwidth]{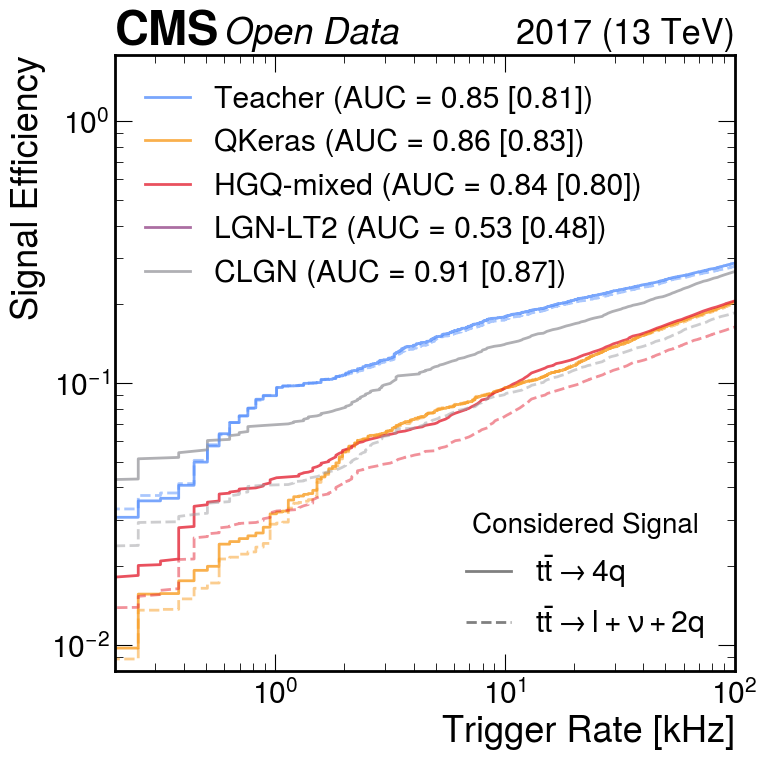}
    \vspace*{-23pt}
    \caption{ROC curves for the two simulated signal samples. The teacher, the baseline student and three selected novel student models are shown. On the x-axis, the trigger rate is shown which is proportional to the false positive rate.}
    \label{fig:roc-curves}
    \vspace{-10pt} % pull text closer after the figure
\end{wrapfigure}

\begin{figure}[htb]
\centering
\includegraphics[width=\textwidth]{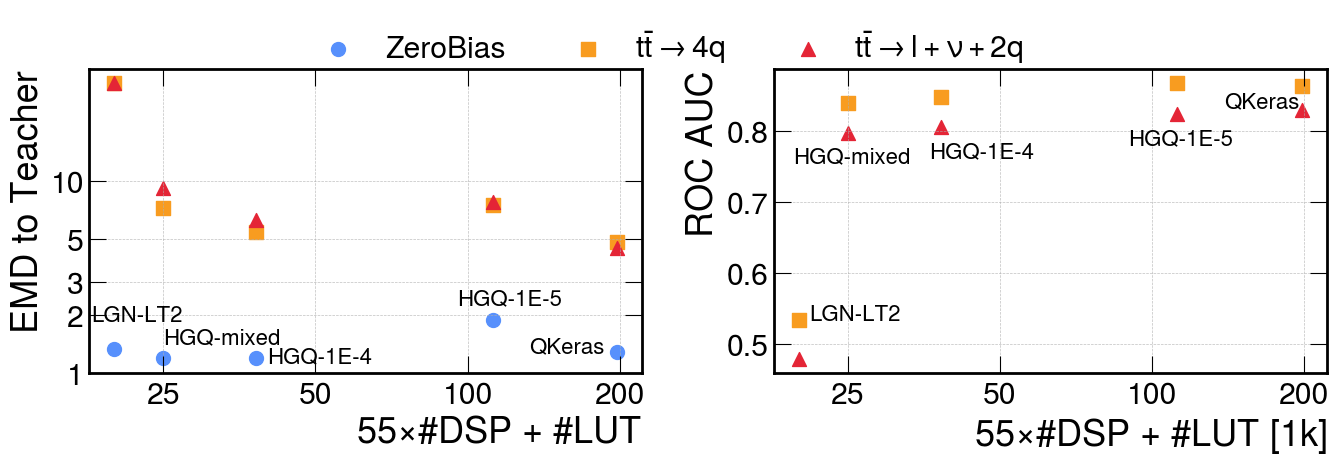}
\caption{Performance results for the different CICADA student models given as a function of resource utilization calculated in EPOPs. Any given model lies along a vertical line. On the left, the EMD between the targets from the teacher and the students' prediction is shown. On the right, the ROC-AUC for two simulated signals is shown. The teacher achieved ROC-AUCs of 0.81 and 0.85 on the $t\bar{t}\rightarrow l+\nu+2q$ and $t\bar{t}\rightarrow4q$ signals, respectively.}
\label{fig:performance_comparison}
\end{figure}

Figure~\ref{fig:roc-curves} shows the classifying performance of selected CICADA models in the form of ROC curves. It can be seen that the convolutional LGN (CLGN), as well as the HGQ-based student with adaptive regularization (HGQ-mixed) achieve comparable performance to the baseline (QKeras). For the most important regime with trigger rates below 1kHz, the novel methods even outperform the baseline. 
Figure~\ref{fig:performance_comparison} (left) shows the EMD as a function of EBOPs (\( \mathrm{EBOPs}\approx \#\mathrm{LUTs}+55*\mathrm{\#DSPs}\)), which was found to be a good estimate for the final resource count, after the place-and-route stage~\cite{HGQ}.
EMD is reported for three datasets: Zero Bias,
$t\bar{t}\rightarrow4q$, and 
$t\bar{t}\rightarrow l+\nu+2q$.
The HGQ student models generally achieve comparable or sometimes lower EMD than the QKeras model, demonstrating strong fidelity in learning the teacher outputs.
The dense LGN student (LGN-LT2) learns the Zero Bias sample reasonably well but struggles with the signal datasets, likely because the limited number of high-value outliers in the training set is insufficient given the binary input representation required. The convolutional LGN overcomes this issue, but its synthesis for FPGA is work-in-progress. The number of logic gates is, however, comparable to the dense architecture shown here.
HGQ models also maintain slightly lower resource usage while achieving similar physics performance, as illustrated in the ROC AUC versus EBOPs plot (right). Table ~\ref{tab:hardware_cost} shows the resource utilization estimates in detail.

\begin{table}[htb]
\centering
\scriptsize
    \def\arraystretch{1}
    \caption{Pre-place-and-route hardware cost comparison for select student models. Latency is reported in clock cycles (cc), where one cycle is 6.25 ns.}
\resizebox{\textwidth}{!}{% Scale to text width

\begin{tabular}{l  cccccc}
    Model Label & Quantization Library & Latency (in cc) & DSPs  & FFs & LUTs & HGQ EBOPs\\
    \hline
    QKeras          & QKeras               & 16 & 697 & 50368 & 159447 & -\\
    HGQ-1E-5        & HGQ                  & 17 & 4 & 27776 & 111848 & 39170\\
    HGQ-1E-4        & HGQ                  & 11 & 1 & 6229 & 38111 & 7570\\
    HGQ-mixed       & HGQ                  & 8 & 0 & 3019 & 24947 & 3301\\
    LGN-LT2         & LGN                  & 3 & 0 & 856 & 19977 & - \\
\end{tabular}
}
\label{tab:hardware_cost}
\end{table}

\section{Limitations and Future work}

The results presented here serve as a proof of concept, and we anticipate that performance can be further enhanced through a detailed exploration of synthesis parameters. 
We are working on further improving the performance of the students using these novel quantization methods. 
Although resource utilization estimates are provided in this paper, performing place-and-route is required to determine the real hardware cost of the student models.
Future work also includes synthesis of convolutional LGNs that demonstrate great accuracy, but for which we do not have resource estimates, yet~\cite{10.5555/3737916.3741767}.

\section{Broader Impact}

By demonstrating a substantial reduction in hardware requirements without compromising physics performance, this work opens the door to deploying more sophisticated machine learning algorithms within the stringent real-time constraints of the CMS L1 Trigger and CICADA. For instance, it now becomes feasible to incorporate additional input features that encode finer details of upstream energy distributions, as well as to enable real-time monitoring for potential input drifts. These advances have the potential to enhance searches for new physics by improving the trade-off between signal efficiency and background rejection.

Beyond the CICADA use case, the demonstrated feasibility of the two novel fast machine learning approaches carries broader significance. Fast ML on resource-constrained hardware is of growing importance for future high-energy physics (HEP) experiments, where tasks such as real-time particle track reconstruction are critical. This applies not only to the HL-LHC, but also to future facilities such as the FCC and potential muon colliders. Traditional optimization techniques—such as fixed-precision quantization and pruning—are unlikely to fully meet the demands imposed by next-generation data rates. Hence, more innovative and radical solutions, like the ones explored here, will be essential. The methods and findings presented in this study thus lay important groundwork for future efforts in fast ML, both within HEP and in other real-time data-intensive domains.
% If the extension to higher-input LUTs proves successful, the methods introduced here could mark the beginning of a new generation of AI models optimized for resource-constrained environments. Such developments have the potential to enable scientific capabilities previously considered infeasible, such as real-time track reconstruction in future collider experiments---thereby substantially expanding the discovery potential of these facilities.

\printbibliography

@misc{petersen2022deepdifferentiablelogicgate,
      title={Deep Differentiable Logic Gate Networks}, 
      author={Felix Petersen and Christian Borgelt and Hilde Kuehne and Oliver Deussen},
      year={2022},
      eprint={2210.08277},
      archivePrefix={arXiv},
      primaryClass={cs.LG},
      url={https://arxiv.org/abs/2210.08277}, 
}

@inproceedings{10.5555/3737916.3741767,
author = {Petersen, Felix and Kuehne, Hilde and Borgelt, Christian and Welzel, Julian and Ermon, Stefano},
title = {Convolutional differentiable logic gate networks},
year = {2025},
isbn = {9798331314385},
publisher = {Curran Associates Inc.},
address = {Red Hook, NY, USA},
booktitle = {Proceedings of the 38th International Conference on Neural Information Processing Systems},
articleno = {3851},
numpages = {19},
location = {Vancouver, BC, Canada},
series = {NIPS '24}
}

@misc{pol2023knowledgedistillationanomalydetection,
      title={Knowledge Distillation for Anomaly Detection}, 
      author={Adrian Alan Pol and Ekaterina Govorkova and Sonja Gronroos and Nadezda Chernyavskaya and Philip Harris and Maurizio Pierini and Isobel Ojalvo and Peter Elmer},
      year={2023},
      eprint={2310.06047},
      archivePrefix={arXiv},
      primaryClass={cs.LG},
      url={https://arxiv.org/abs/2310.06047}, 
}

@misc{HGQ,
  doi = {10.7907/HQ8JD-RHG30},
  url = {https://authors.library.caltech.edu/doi/10.7907/hq8jd-rhg30},
  author = {Chang, Sun and Årrestad, Thea and Lončar, Vladimir and Ngadiuba, Jennifer and Spiropulu, Maria},
  language = {en},
  title = {Gradient-based Automatic Per-Weight Mixed Precision Quantization for Neural Networks On-Chip},
  publisher = {California Institute of Technology},
  year = {2024},
  copyright = {No commercial reproduction, distribution, display or performance rights in this work are provided.}
}

@techreport{CMS-DP-2024-121,
  author       = {{CMS Collaboration}},
  title        = {{Model-Independent Real-Time Anomaly Detection at the CMS Level-1 Calorimeter Trigger with CICADA}},
  type         = {CMS Detector Performance Summary},
  institution  = {CERN},
  number       = {CMS-DP-2024-121},
  month        = nov,
  year         = {2024},
  note         = {Available at: \url{https://cds.cern.ch/record/2917884}},
}

@techreport{CMS-DP-2024-059,
  author       = {{CMS Collaboration}},
  title        = {{2024 Data Collected with AXOL1TL Anomaly Detection at the CMS Level-1 Trigger}},
  type         = {CMS Detector Performance Summary},
  institution  = {CERN},
  number       = {CMS-DP-2024-059},
  month        = nov,
  year         = {2024},
  note         = {Available at: \url{https://cds.cern.ch/record/2904695}},
}

@manual{XilinxVivado2021,
  title = {{Xilinx Vivado Design Suite User Guide: Release Notes, Installation, and Licensing}},
  author = {{Xilinx Inc.}},
  year = {2021},
  note = {Version 2021.1},
  url = {https://www.xilinx.com/support/documentation/sw_manuals/xilinx2021_1/ug973-vivado-release-notes-install-license.pdf},
  institution = {Xilinx Inc.},
  month = jun,
  address = {San Jose, CA}
}

@manual{XilinxVitisHLS2021,
  title = {{Xilinx Vitis High-Level Synthesis User Guide}},
  author = {{Xilinx Inc.}},
  year = {2021},
  note = {Version 2021.1},
  url = {https://www.xilinx.com/support/documentation/sw_manuals/xilinx2021_1/ug1399-vitis-hls.pdf},
  institution = {Xilinx Inc.},
  month = jun,
  address = {San Jose, CA}
}

@article{qkeras,
	author = {Coelho, Claudionor N. and Kuusela, Aki and Li, Shan and Zhuang, Hao and Ngadiuba, Jennifer and Aarrestad, Thea Klaeboe and Loncar, Vladimir and Pierini, Maurizio and Pol, Adrian Alan and Summers, Sioni},
	journal = {Nature Machine Intelligence},
	number = {8},
	pages = {675--686},
	title = {Automatic heterogeneous quantization of deep neural networks for low-latency inference on the edge for particle detectors},
	volume = {3},
	year = {2021}}

@article{hlsfml,
      author         = "Duarte, Javier and others",
      title          = "{Fast inference of deep neural networks in FPGAs for
                        particle physics}",
      journal        = "JINST",
      volume         = "13",
      year           = "2018",
      number         = "07",
      pages          = "P07027",
      doi            = "10.1088/1748-0221/13/07/P07027",
      eprint         = "1804.06913",
      archivePrefix  = "arXiv",
      primaryClass   = "physics.ins-det",
      reportNumber   = "FERMILAB-PUB-18-089-E",
      SLACcitation   = "%%CITATION = ARXIV:1804.06913;%%"
}

@book{trigger_tdr,
      collaboration = "CMS",
      title         = "{CMS TriDAS project: Technical Design Report, Volume 1:
                       The Trigger Systems}",
      series        = "Technical design report. CMS",
      url           = "https://cds.cern.ch/record/706847",
}

@techreport{trigger_upgrade_tdr,
      author        = "Tapper, A and Acosta, Darin",
      collaboration = "CMS",
      editor        = "Tapper, A",
      title         = "{CMS Technical Design Report for the Level-1 Trigger
                       Upgrade}",
      reportNumber  = "CERN-LHCC-2013-011, CMS-TDR-12",
      year          = "2013",
      url           = "https://cds.cern.ch/record/1556311",
      note          = "Additional contacts: Jeffrey Spalding, Fermilab,
                       Jeffrey.Spalding@cern.ch Didier Contardo, Universite Claude
                       Bernard-Lyon I, didier.claude.contardo@cern.ch",
}

@misc{CERNOpenData2017,
  title        = {CERN Open Data Portal},
  author = {{CMS Data preservation and open access group}},
  year         = {2025},
  howpublished = {\url{https://opendata.cern.ch/}},
author = {{CMS Data preservation and open access group}},
  note         = {Accessed: 2025-08-30}
}
% \section{Potentially useful papers to cite}
% Deep Differentiable Logic Gate Networks\cite{10.5555/3600270.3600416}\\
% Convolutional LGNs\cite{10.5555/3737916.3741767}\\
% Highly Granular Quantization
% \cite{HGQ} \todo{accepted at ICML, waiting for proper citation}\\
% Logic Nets \cite{umuroglu2020logicnetscodesignedneuralnetworks}\\
% Adrian's AD w/ Knowledge Distillation\cite{pol2023knowledgedistillationanomalydetection}\\
% Binary NN survey \cite{Qin_2020}\\
% QKeras\cite{Coelho_2021} [check]

%\bibliographystyle{plain}
%\bibliography{mybib}
% \pagebreak
\appendix
\section{Existing CICADA Architectures}
\begin{figure}[htb]
    \centering
    % \begin{subfigure}[b]{0.6\textwidth}
        % \centering
        \includegraphics[width=0.6\textwidth]{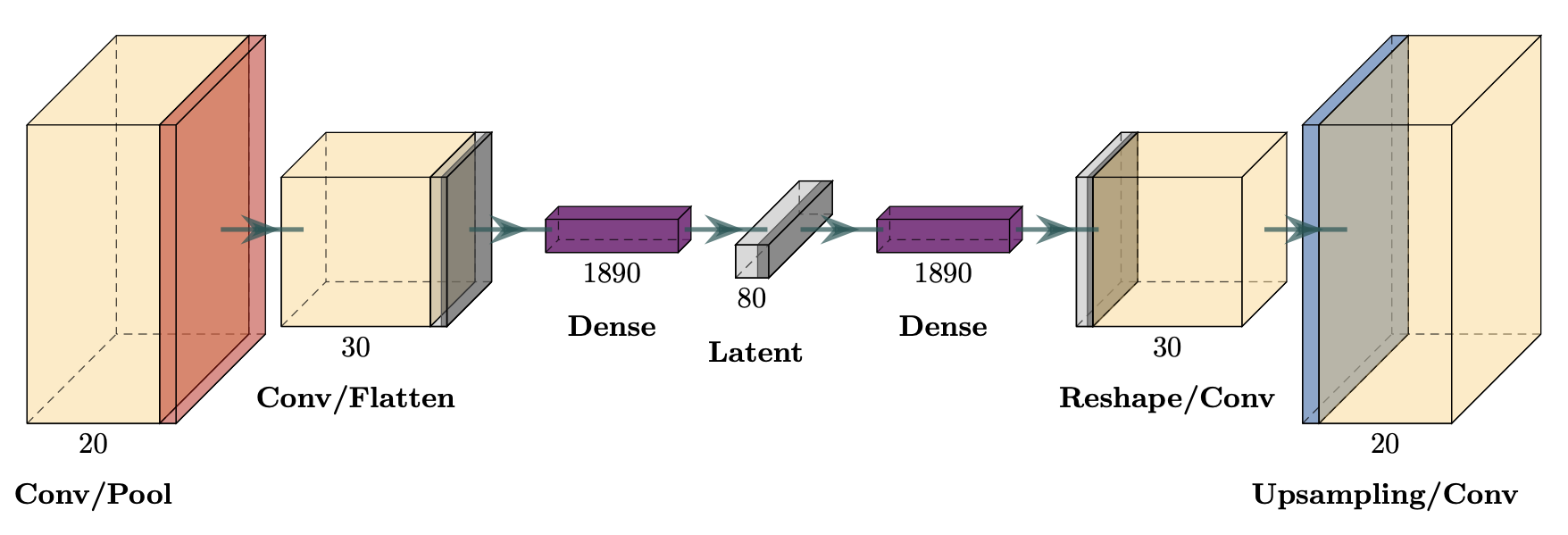}
        \caption{CICADA teacher model architecture.}
        \label{fig:teacher_arch}
    % \end{subfigure}
    % \hfill 
    % \begin{subfigure}[b]{0.3\textwidth}
        % \centering
        \includegraphics[width=0.3\textwidth]{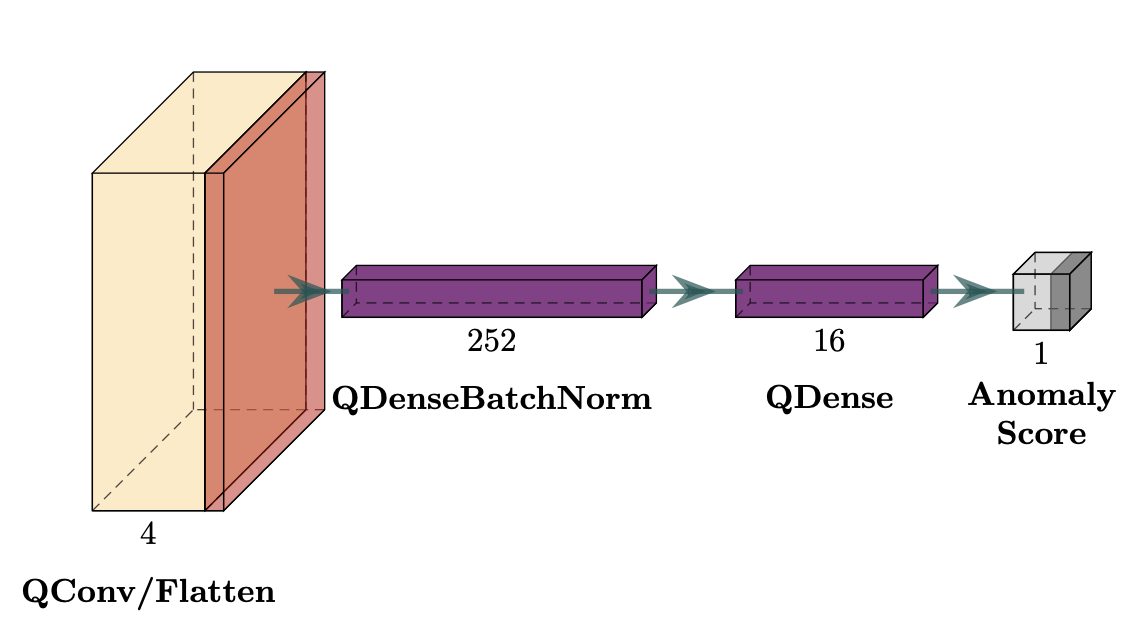}
        \caption{CICADA legacy student model architecture.}
        \label{fig:student_arch} 
    % \end{subfigure}
    \caption{The numbers at the convolutional layers represent the number of filters used in each layer (e.g., 20, 30), while the numbers at the dense layers indicate the number of nodes in each fully connected layer (e.g., 1890, 80).}
    \label{fig:model_arch}
\end{figure}

The teacher model for CICADA is a convolutional autoencoder that follows the architecture outlined in Fig. ~\ref{fig:teacher_arch}. 
The encoding step comprises two convolutional layers followed by a dense layer, which compresses the input into the low-dimensional latent space.
The decoding step follows these steps in reverse.
The currently implemented student model on which this paper attempts to improve is a much smaller convolutional autoencoder, quantized using QKeras.
The student model follows the architecture outlined in Fig. ~\ref{fig:student_arch}.

\section{Data Representation for LGN Models}
During training, the LGN architecture can handle float inputs, since each logic gate is represented as a continuous function. 
However, the compiled model uses discrete logic gates in place of the continuous functions and can therefore only accept binary inputs.
The schema for representing the $18\phi\times14\eta$ integer input grid in binary form during training and for the compiled model has a large effect on the quality of anomaly score reconstruction.
This allows values between 0 and 1 while training, such that when the inputs are collapsed to binary values upon compiling the model, important information is retained.\\
The bit representation $\text{lt}2$ follow a log transformation function. Let $x$ be the input and $\{t_i\}_{i=1}^{2}$ be a list of thresholds. 
Then the transformation is calculated as:
\begin{equation}
    y_i = \log\Big(1 + \frac{\max(x - t_i, 0)}{2}\Big), \quad i = 1, \dots, 2
\end{equation}
Different thresholds and number of thresholds are selected for different student models.
The untransformed inputs range from 0 to 256, but high values are extremely rare in the input dataset.
The threshold values are therefore chosen as [0, 2] to reflect this.
%\section{Training Details}
%\subsection{HGQ Training Details}
%\subsection{LGN Training Details}

\section{Source Code}
The source code to reproduce the results can be found on GitHub: \href{https://anonymous.4open.science/r/ml4physics-paper-plots-7C16/}{https://anonymous.4open.science/r/ml4physics-paper-plots-7C16/}. Our implementation of logic gate networks is publicly available on GitHub: \href{https://github.com/ligerlac/torchlogix}{https://github.com/ligerlac/torchlogix}
\end{document}